# Electromechanically Tunable Metasurface Transmission Waveplate at Terahertz Frequencies

Xiaoguang Zhao[1], Jacob Schalch[2], Jingdi Zhang[2], Huseyin R. Seren[1], Guangwu Duan[1], Richard D. Averitt[2,*], and Xin Zhang[1,*]

1. Department of Mechanical Engineering, Boston University, Boston, Massachusetts 02215, USA.

2. Department of Physics, University of California, San Diego, La Jolla, California 92093, USA.

*Authors to whom correspondence should be addressed: RDA (email: raveritt@ucsd.edu) and XZ (email: xinz@bu.edu)

**Abstract**

Dynamic polarization control of light is essential for numerous applications ranging from enhanced imaging to materials characterization and identification. We present a reconfigurable terahertz metasurface quarter-waveplate consisting of electromechanically actuated micro-cantilever arrays. Our anisotropic metasurface enables tunable polarization conversion cantilever actuation. Specifically, voltage-based actuation provides mode selective control of the resonance frequency, enabling real-time tuning of the polarization state of the transmitted light. The polarization tunable metasurface has been fabricated using surface micromachining and characterized using terahertz time domain spectroscopy. We observe a ~230 GHz cantilever actuated frequency shift of the resonance mode, sufficient to modulate the transmitted wave from pure circular polarization to linear polarization. Our CMOS-compatible tunable quarter-waveplate enriches the library of terahertz optical components, thereby facilitating practical applications of terahertz technologies.

**Introduction**

Precise control of the polarization state of an electromagnetic wave is of great importance in applications in astrophysics, chemistry, microscopy and advanced optics[1]. Conventionally, material birefringence is used to construct polarization sensitive optical components such as quarter-wave and half-wave plates[2]. However, the majority of polarization components are limited by the material response and lack of tunability. A notable exception is commercially available photo-elastic modulators that efficiently tune the polarization state using piezoelectric transducers, though they are relatively bulky and working at mid-infrared and shorter wavelength regimes. The development of artificial electromagnetic (EM) materials, including metamaterials and metasurfaces, provides a route to break the limitation of natural materials, enabling versatile and compact control of EM wave propagation from microwave to visible frequencies[4,5]. Design based control of the effective permittivity and permeability achieved with metamaterials has led to the realization of negative refractive index[6] and controllable chirality[7,8] and enabled, as examples, superlensing[9], perfect absorption[10], and transformation optics for advanced wave manipulation[11-13]. Based on the generalized laws of refraction and reflection[14], planar metasurfaces can control the phase front of the outgoing wave for beam focusing[15], steering[16], amongst others[17-19]. In particular, anisotropic metamaterials and metasurfaces support polarization conversion[20-24] and optical activity[25-28].

With metamaterials or metasurfaces, it is possible to not only obtain the desired permittivity, permeability, and chirality at a chosen frequency, but also tune and reconfigure the effective parameters to construct so-called "metadevices"[29]. Different approaches, including optical excitation[30-33], electrical gating[34-36], phase change[37,38], and mechanical actuation[39-48] have been investigated to modulate the amplitude and phase response of metamaterials. Notable polarization based devices include a switchable quarter-waveplate[49] and tunable optical activity[42] to extend the operating bandwidth and modulate the polarization state. Recently, a metamaterial device consisting of microelectromechanical systems (MEMS) bimorph cantilevers has been demonstrated to control the polarization of terahertz radiation[50]. This device operates in reflection, complementing our transmission-based device as discussed in greater detail below.

In this paper, we employ single-layer micro-cantilevers to construct a terahertz (THz) metasurface with a highly anisotropic response, allowing us to tune the resonance frequency for radiation polarized along one axis (x-axis) without affecting the response along the orthogonal direction (y-axis). The large resonance frequency tuning range (~230 GHz) achieved with electrostatically actuated cantilevers modifies the polarization of the transmitted wave from circular polarization to linear polarization, yielding a reconfigurable quarter-waveplate operating in transmission. Our theoretical analysis based on an equivalent circuit model and finite element simulations unveil the physics of the experimental results. The dimensions of the structure can be adapted to design metasurface devices for other frequency ranges, including microwave and infrared. Our device is fabricated using surface micromachining, which is compatible with CMOS, providing opportunities for reconfigurable and reprogrammable quarter-waveplates and other polarization control devices.

**Materials and Methods**

The working principle of the tunable metasurface is shown in Fig. 1. The unit-cell is a suspended cantilever, of which the free-end overlaps with an underlying capacitive pad. For x-polarized normal incidence light ($\theta = 0°$ in Fig. 1a), the unit-cell can be considered as a second order LC resonator (as shown in Fig. 1b), in which the inductance is associated with the cantilever and the capacitance is formed by the cantilever tip and underlying pad. The array of unit-cells is patterned on a slightly doped silicon wafer that is coated with an insulating silicon nitride thin film. A voltage across the cantilevers and the substrate induces an electrostatic force to pull the cantilevers downward. The change in capacitance modifies the resonant response. Conversely, for the y-polarized incident wave ($\theta$ = 90° in Fig. 1a), only the wires connecting adjacent cantilevers (which can be modeled as an inductor, as shown in Fig. 1c) need to be considered. The overall transmission response of the metasurface, which is represented by the Jones vector and transmission matrix **T**, can be written as[1]

$$\begin{bmatrix} E_x^t \\ E_y^t \end{bmatrix} = \mathbf{T} \begin{bmatrix} E_x^i \\ E_y^i \end{bmatrix} = \begin{bmatrix} t_{xx} & t_{xy} \\ t_{yx} & t_{yy} \end{bmatrix} \begin{bmatrix} E_x^i \\ E_y^i \end{bmatrix} \quad (1)$$

where $E_x^i$ and $E_y^i$ are the x- and y-polarized components of the incident wave, respectively; $E_x^t$ and $E_y^t$ denote the transmitted electric field along x- and y-axis, respectively. For the transmission matrix of the cantilever metasurface, the off-diagonal elements are negligible due to weak cross-polarization coupling. The diagonal element complex transmission coefficients $t_{xx}$ and $t_{yy}$, can be modeled by the following equations according to transmission line theory[51]:

$$t_{xx} = \frac{2Z_{xx}}{Z_{xx} + Z_0} \cdot D \quad (2)$$

$$t_{yy} = \frac{2Z_{yy}}{Z_{yy} + Z_0} \cdot D \quad (3)$$

where $Z_{xx}$ and $Z_{yy}$ are effective impedances for the x- and y-polarization directions, respectively, $Z_0$ is the impedance of free space, and $D$ represents the transmission loss at the substrate/air interface. According to the equivalent circuit model shown in Fig. 1b and 1c, we can express these impedances as $Z_{xx} = R_1 + i\omega L_1 + 1/(i\omega C_1)$ and $Z_{yy} = R_2 + i\omega L_2$. Simply speaking, cantilever actuation modifies the impedance $Z_{xx}$ by changing the capacitance $C_1$ without affecting $Z_{yy}$, leading to modulation of the transmission characteristics for x-polarized incidence. As such, the anisotropy of the cantilever array and associated response enables control of the polarization state of the transmitted wave as discussed below.

The designed structure was fabricated using surface micromachining on a slightly doped silicon substrate. The fabrication process flow is shown in Fig. S1. The fabricated tunable metasurface and the mechanical response are shown in Fig. 2. A DC voltage was applied to the cantilever and the substrate through the bonding wires (Fig. 2a). The area of the cantilever array is 8×8 mm$^2$. The rectangular hole in the beam is designed for ease of releasing and the dimple in the free-end of the beam enhances the capacitance change and prevents adhesion between the cantilever beam and the underlying capacitive pad (Fig. 2b). The cantilevers bend upwards due to residual stress after releasing. When a voltage is applied, the electrostatic force between the cantilever and the substrate pulls the cantilevers downwards. The beam curvature as a function of applied DC voltage has been characterized using a laser interferometer (ZYGO), as shown in Fig. 2c. Initially, the height of the tip, i.e. the distance between the bottom of the dimple and the top of the capacitive pad, is ~0.9 μm, as shown in Fig. 2d. With increasing voltage, the beam was pulled downward by the electrostatic force between the cantilever and the substrate. The pull-in voltage, at which the cantilever snaps down to the capacitive pad, is approximately 38 V.

The tunable transmission of the metasurface was measured using terahertz time domain spectroscopy (THz-TDS), as shown in Fig. S2 in the supplemental information. An 800 nm Ti:Sapphire laser with 25 fs pulse duration and 80 MHz repetition rate was employed to excite a biased photoconductive antenna to generate THz pulses. THz pulses were focused on the metasurface sample and transmitted pulses were sampled by another photoconductive antenna to measure the time resolved transmission signal, which was Fourier transformed to the frequency domain ($E_{sample}(\omega)$). The spectrum of the incident pulse, denoted by $E_{ref}(\omega)$, was measured using air as the reference. The transmission spectrum of the metasurface can be calculated by $t(\omega) = E_{sample}(\omega)/E_{ref}(\omega)$. Since the response of the metasurface was anisotropic, we measured the transmission spectra for two orthogonal polarization states, i.e. $t_{xx}$ and $t_{yy}$, by rotating the sample by 90°. A DC voltage source was applied to actuate the cantilevers and the transmission coefficients ($t_{xx}$ and $t_{yy}$) were measured for a given voltage. Time domain signals were windowed to remove etalon artifacts as described in the supplemental information.

From the measured transmission coefficients, the x- and y- polarized components of the transmitted waves can be calculated with the Jones matrix [Eq. (1)]. Linearly polarized normal incidence radiation is described by $\vec{E}^i(\omega) = (E_x^i \vec{x} + E_y^i \vec{y})e^{i(-\omega z/c + \omega t)}$, in which $\vec{x}$ and $\vec{y}$ denote the unit vector along x- and y- axes, respectively, $\omega$ is the angular frequency, c is the speed of light, and z is the position. Given a polarization angle $\theta$ (as shown in Fig. 1a), we can obtain $E_x^i = |\vec{E}^i(\omega)|\cos\theta$ and $E_y^i = |\vec{E}^i(\omega)|\sin\theta$. The x and y components of the transmitted wave are $E_x^t = t_{xx}E_x^i$ and $E_y^t = t_{yy}E_y^i$, for which $t_{xx}$ and $t_{yy}$ are the transmission coefficients for x- and y-polarization, respectively. The polarization information of the transmitted wave can be represented by the Stokes parameters[1] [$S_0$, $S_1$, $S_2$, $S_3$], in which $S_0 = |E_x^t|^2 + |E_y^t|^2$, $S_1 = |E_x^t|^2 - |E_y^t|^2$, $S_2 = 2Re(E_x^t E_y^{t*})$, and $S_3 = 2Im(E_x^t E_y^{t*})$. Among the parameters, $S_0$ (= $I$) is the total intensity of the wave, $S_1$ (= $Ip\cos(2\psi)\cos(2\chi)$) is the component of horizontally or vertically linear polarization, $S_2$ (= $Ip\sin(2\psi)\cos(2\chi)$) is the component of 45° or -45° linear polarization, and $S_3$ (= $Ip\sin(2\chi)$) is the component of circular polarization, $p$ is the degree of polarization, $\psi$ is the orientation angle, and $\chi$ is the elliptical angle, as shown in Fig. S5b in the supplemental information. The circular polarization ratio (*CPR* = $S_3/S_0$) and axial ratio ($AR = \tan(\chi)$) can be employed to describe the polarization state of a wave.

A 3D model of the metasurface was created in CST Microwave Studio. In the model, a unit-cell cantilever with corresponding capacitive pad on the silicon substrate was included and periodic boundary conditions were employed. The silicon substrate was treated as a lossy dielectric material with a permittivity of 11.86 and a conductivity of 0.1 S/cm, silicon nitride was considered as a dielectric material with permittivity of 7.0, gold is metallic with conductivity of $4.6 \times 10^5$ S/cm, and copper is metallic with conductivity of $5.8 \times 10^5$ S/cm. Since the silicon wafer we used is slightly doped with nominal conductivity of 0.1 S/cm, the transmission is attenuated by ~ 40% when the THz pulses propagate through the substrate according to our simulation, which is validated by transmission coefficient in the experimental results shown below. The geometric parameters in the simulations are consistent with those of the fabricated sample. The frequency solver was utilized to simulate the transmission coefficients for both polarizations. For the initial case ($V_{DC}$ = 0), the measured curvature of the cantilever, as shown in Fig. 2c, was applied. The tip height of the cantilever was adjusted to match the measured results for different applied voltages.

**Results and Discussion**

The transmission spectra of the metasurface for different applied voltages has been characterized using terahertz time domain spectroscopy (THz-TDS), as described above. To verify the metasurface anisotropy, the transmission for x- and y-polarized terahertz pulses (Figs. 3a-3d) were measured by rotating the sample 90º about the incident axis. At zero voltage, a strong resonance is observed for x-polarization at 1.04 THz with a transmission amplitude of 0.03, while there is no obvious resonance observed for y-polarization. Numerical simulations were performed using CST Microwave Studio (details in the Materials and Methods section) to study the metasurface EM response. The simulations show good agreement with experimental results of the transmitted amplitude and phase. For x-polarization at 1.04 THz, currents are excited in the copper cantilever structure, corresponding to the LC resonance mode as illustrated in Figs. 3e and 3f. The electric field (Fig. 3f) is concentrated in the gap between the cantilever tip and the underlying capacitive pad, consistent with the tip-pad structure being the dominant contributor to the overall capacitance. However, the fringe field along the cantilever beam indicates that the contribution to the capacitance between the beam and silicon substrate cannot be ignored. For y-polarized THz pulses, current in the connection wires (Fig. 3g) are present, equivalent to an inductor as illustrated in Fig. 1c with, as expected, no electric resonance (Fig. 3h). Quantitatively, the equivalent circuit model [Eqs. (2) and (3)] is validated by fitting the experimental and simulation results with proper parameters, including $R_1$, $L_1$, $C_1$, $R_2$, $L_2$, and $D$, in the circuit model, as shown in Figs. S3 and S4 in the supplemental information.

With an applied voltage, the cantilevers are pulled downward and the resonance frequency of the metasurface redshifts for the x-polarized THz pulses due to the increased capacitance, as shown in Fig. 4a. The maximum resonance frequency shift is ~230 GHz, at 40V DC bias. The frequency shift as a function of applied voltage saturates for voltages > 40 V since the cantilevers are pulled down to the capacitive pads, preventing further deformation of the beam. The amplitude of $t_{xx}$ is strongly modified at certain frequencies. For example, at 0.81 THz and 1.04 THz absolute modulation of the transmission amplitudes of 34% and 26% are achieved, respectively, as shown in Fig. 4b. The resonance fully recovers to the original zero voltage state when the voltage is turned off. The cantilever-pad spacing was modified in the numerical simulations to match experimental results, indicating that the tunable transmission coefficient originates from deformation of the cantilever. The tunability for x-polarized THz pulses demonstrates that the metasurface operates as a THz modulator[44-46,51].

In contrast, for y-polarized ($t_{yy}$) THz pulses, the transmission amplitude and phase are unchanged upon deflection of the cantilevers (and therefore the applied voltage) as shown by red curves in Figs. 3a-3d. In short, the results in Fig. 3 demonstrate the anisotropic tunability of the cantilever metasurface, which enables creating a tunable quarter-waveplate as we now discuss.

Based on the Jones vector and measured transmission spectra ($t_{xx}$ and $t_{yy}$), we can calculate the Stokes parameters[1], including $S_0$, $S_1$, $S_2$, and $S_3$, which fully describe the polarization state of a transmitted wave for an arbitrary incident polarization, as detailed in the Materials and Methods section.

We consider a normally incident wave with a polarization angle ($\theta$ in Fig. 1a) of 34°. The full set of Stokes parameters can be presented by the Poincare plot (Fig. S5 in the supplementary information). For clarity, we use some key parameters derived from Stokes parameters, including the total intensity (*I*), axial ratio (*AR*), and circular polarization ratio (*CPR*), to describe the characteristics of the transmitted wave. *I* is defined as the ratio of the transmitted intensity to the incident intensity, *AR* is the ratio of the minor to major axes of the transmitted polarization ellipse, and *CPR* is the ratio of the intensity of the circularly polarized wave to the overall transmitted intensity[20]. Ideally, |*AR*| and |*CPR*| are 1 for a pure circularly polarized wave while they are 0 for a pure linearly polarized wave. Of note, the sign of CPR represents the handedness of the circularly polarized wave: a positive sign corresponds to left-hand polarization while a negative sign to a right-hand polarization. Experimentally measured *AR*s and *CPR*s, as shown in Figs. 5a and 5b, reveal voltage control of the polarization state through cantilever deflection. When $V_{DC}$ is 0 V, *AR* and *CPR* of the transmitted wave are close to 0 at the frequency of 1.05 THz indicative of a linearly polarized wave. At 0.81 THz, *AR* ≈ 0.91 and *CPR* ≈ -1, indicating a right-hand circularly polarized wave. In the frequency span from 0.792 THz to 0.842 THz, AR is higher than 0.85 and CPR smaller than -0.99, yielding a circular polarization bandwidth of ~50 GHz. The intensity is frequency-dependent, as shown in Fig. 5c, and obtains minimum values at the resonance frequencies at which the x-polarized wave is filtered out in the transmission. Qualitatively, the polarization state of the transmission can be explained by the transmission spectra shown in Fig. 3a. At 1.05 THz, $|t_{xx}|$ and $|t_{yy}|$ are 0.04 and 0.57, respectively, with a 2º phase difference, which indicates a linear polarization. On the other hand, $|t_{xx}|$ and $|t_{yy}|$ are 0.34 and 0.52, respectively, with an 85º phase difference at 0.81 THz, corresponding to a transmitted beam with nearly circular polarization.

Due to the resonance frequency shift of $t_{xx}$ induced by cantilever actuation, the AR and CPR spectra exhibit a significant shift to lower frequencies. The frequencies for pure circular and linear polarized radiation redshift gradually as $V_{DC}$ increases as illustrated in Figs. 5a and 5b. The shifting of AR and CPR leads to modulation of the polarization state over the whole frequency range. In particular, AR is tuned from 0.91 to 0.01 and CPR from -1 to 0 at 0.81 THz as $V_{DC}$ increases from 0 V to 40 V, showing that the polarization is switched from circular to linear state, as shown in Fig. 5d. Top column in Fig. 5d demonstrates polarization states at different voltages using the electric field components along the x- and y- axes. When $V_{DC}$ is 0 V, the transmission is circular. As $V_{DC}$ increased to 25 V, the transmission is elliptical polarized with orientation angle of 93º, and AR of 0.3 corresponding to ellipticity angle of 32º. At $V_{DC}$ = 40 V (exceeding the pull-in voltage), the linear polarized wave is transmitted from the metasurface with orientation angle of 86º. The amplitude of the transmitted waves is almost constant along the major axis. Examples of tunable polarization at other frequencies are presented in the supplemental information (Fig. S6).

From Fig. 5c, we can observe that our device exhibits a fair amount of insertion loss due mainly to reflection at the air/metasurface interface and limited polarization conversion efficiency. In theory, the maximum transmitted intensity is 50% for a single layer metamaterial waveplate [21], while our tunable metasurface exhibits a transmittance of ~20% for circular polarization (e.g. $V_{DC}$ = 0V, 0.81 THz) and ~10% for linear polarization (e.g. $V_{DC}$ = 40V, 0.81 THz). Even though the transmitted intensity is not ideal, our device achieves a significant and functional tunability of the polarization state. Moreover, the insertion loss can be decreased by adjusting the geometry of the quarter-waveplate to improve the conversion efficiency and eliminate reflective losses. Specifically, optimizing the anisotropy of the metasurface will increase the polarization conversion efficiency[19], and adding a metasurface anti-reflection layer will eliminate the reflection loss[21]. In short, a high-efficiency waveplate with a large tunable response is expected by further optimizing the metasurface.

The tunability of the polarization state exists not only for an incident polarization angle of 34º, but also for arbitrary polarization angles. The calculated CPRs from the measured Stokes parameters for different incident polarization angles ($\theta$) from 0º to 180º show the polarization state of transmitted waves, presented in Figs. 5e-5g. When $V_{DC}$ = 0, right-hand circular polarized transmission can be achieved for incident waves with polarization angles from 0º to 57º, while left-hand circular polarization is achievable for polarization angles from 123º to 180º over this frequency range (0.4 – 1.0 THz). As the applied voltage increases, it seems that the overall spectra of CPR shift to lower frequencies, demonstrating the tunable polarization state of transmitted waves for arbitrarily polarized incidence.

Compared with the MEMS-cantilever based tunable polarization control recently demonstrated [50], the design and performance of our device is different in several respects. First, our device is designed to operate in transmission whereas in Ref. [50] the device operated in reflection mode. In Ref. [50] actuation of bimorph cantilevers is used to engineer the losses, leading to significant modulation of the resonance frequency amplitude in addition to a frequency shift. It exhibits large tuning range of the phase response and polarization conversion among right-handed circular polarization, liner polarization and left-handed circular polarization. This requires accurate voltage control for polarization conversion. For our device, copper cantilevers modulate the effective capacitance in the equivalent RLC circuit model of the metasurface, which shifts the resonance frequency without a large effect on the resonant transmission amplitude. While our metasurface can operate at intermediate voltages, it can easily switch between quarter waveplate and half waveplate by simply increasing the voltage from 0 V to 40 V (pull-in) voltage. In addition, the fabrication the metamaterial in Ref. [50] requires multiple steps of atomic layer deposition of aluminum oxide, which enables better insulation. In contrast, the fabrication process of our device is less involved and with better yield. Our present device does have a fairly large insertion loss, which can be reduced with additional engineering. In short, the two devices are complementary and provide new insights into MEMS based polarization control of THz radiation.

**Conclusions**

We have demonstrated a cantilever array based metasurface to control the characteristics of transmitted THz fields, including amplitude, phase and polarization. The metasurface is designed with an anisotropic response where electromechanical actuation of the cantilevers shifted the resonance frequency for x-polarization by ~230 GHz with an applied voltage of 40 V without affecting the y-polarization. We are able to tune the polarization of transmitted light from circular to linear at 0.81 THz. The significant modulation of the polarization can be achieved at other frequencies by modifying the geometrical parameters of the metasurface. The ability of microelectromechanical cantilever actuators is not limited to polarization control (as demonstrated above), and they can be integrated with other metasurface and metamaterial configurations for different functions[44-46,52]. In particular, cantilevers provide a route to control the response of each unit-cell individually with well-designed routing strategies, similar to deformable mirrors[53]. For example, we can combine microcantilever actuators with gradient metasurface structures that are designed for beam steering or focusing[54] to enable real-time tunable devices by controlling the phase discontinuities at the unit-cell level[46,50,52], as well as other applications, such as digital coding metasurfaces[55-58]. This paper presents a micro-cantilever based reconfigurable quarter-waveplate to manipulate the polarization of THz radiations dynamically, facilitating the development and applications of THz technologies.

**Acknowledgements**

The authors acknowledge the National Science Foundation under Grant No. ECCS-1309835 and ARO W911NF-16-1-0361. We would like to thank Boston University Photonics Center for technical support.

**Conflict of interests**

The authors declare no conflict of interests.

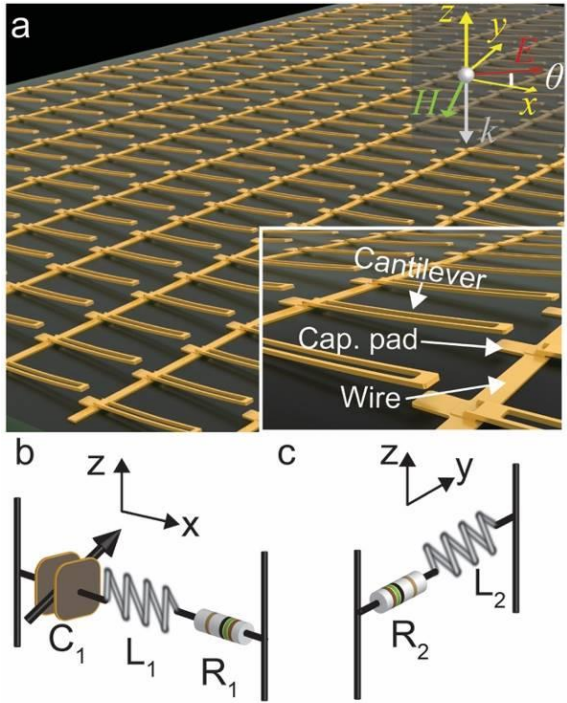

Figure 1. (a) The schematic of the tunable cantilever metasurface array. Inset: close-up view of the metamaterial. Each unit-cell consists of a cantilever, capacitive pad and interconnect wire. The electromagnetic wave is normally incident with a given polarization angle $\theta$. (b), (c) Equivalent circuit models for the x ($\theta = 0°$) and y ($\theta = 90°$) polarizations, respectively.

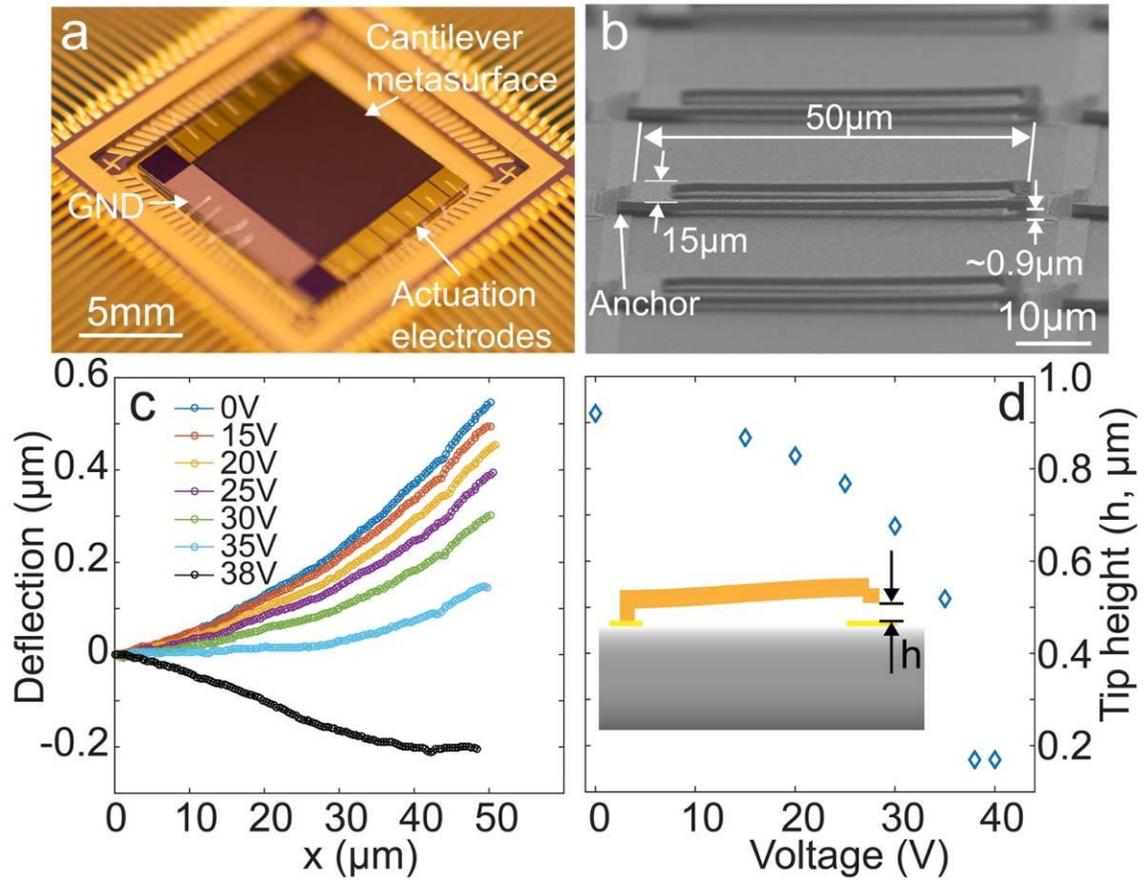

Figure 2. (a) Photograph of an integrated metasurface chip and (b) scanning electron microscope (SEM) image of the metasurface. (c) Deflection profile of the cantilever curvature at different voltages. (d). The measured tip height versus applied DC voltage, with a pull-in voltage of ~38 V.

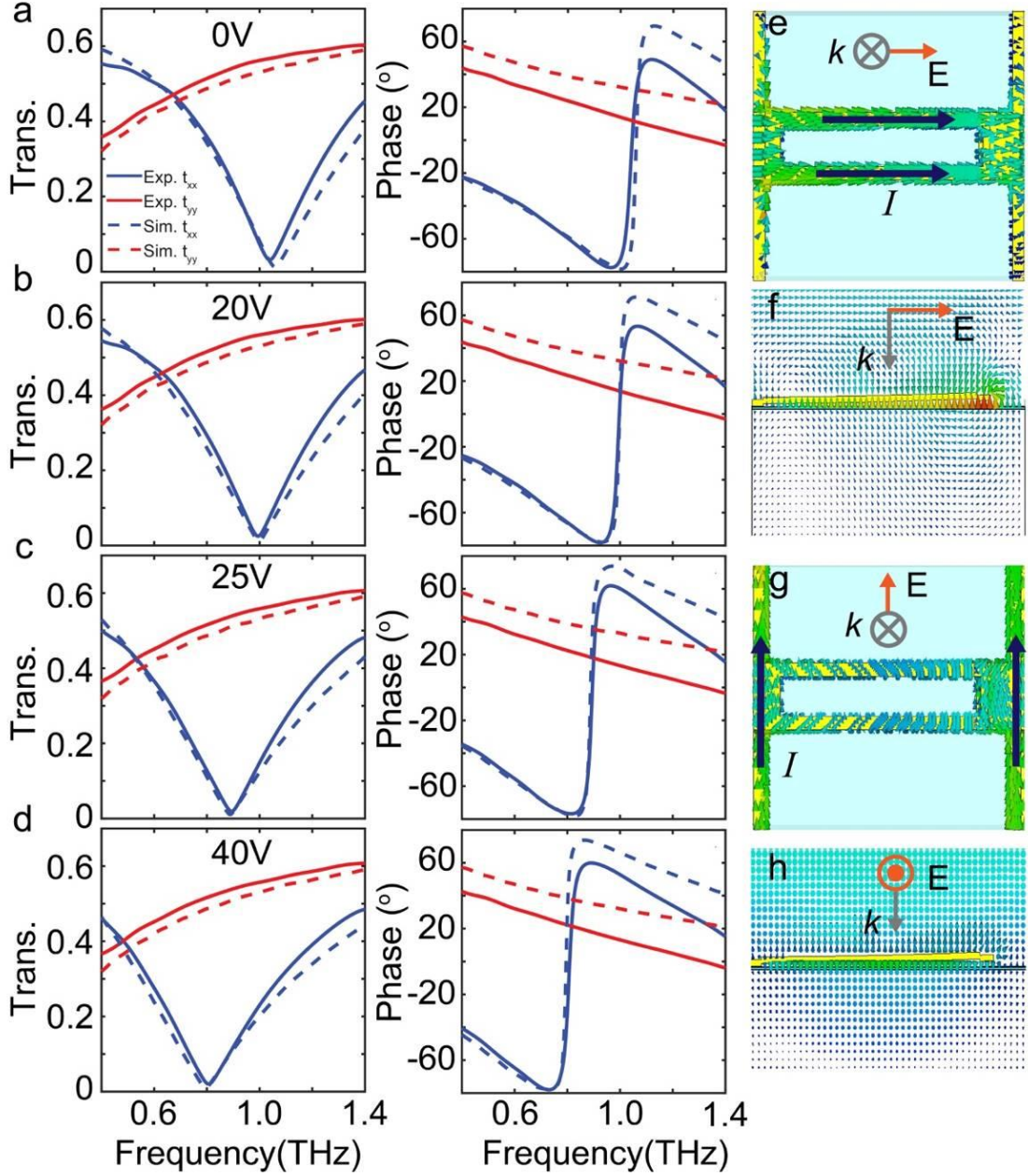

Figure 3. (a) – (d) Amplitude (left) and phase (right) of the transmission coefficients for x ($t_{xx}$, blue lines) and y ($t_{yy}$, red lines) polarization at different voltages. The solid lines are experimental results while the dashed lines are from simulation. (e), (f) On resonance simulated current and electric field distribution, respectively, for x-polarization, and (g) and (h) are for the y-polarization (note: (e) – (h) are for 0 V).

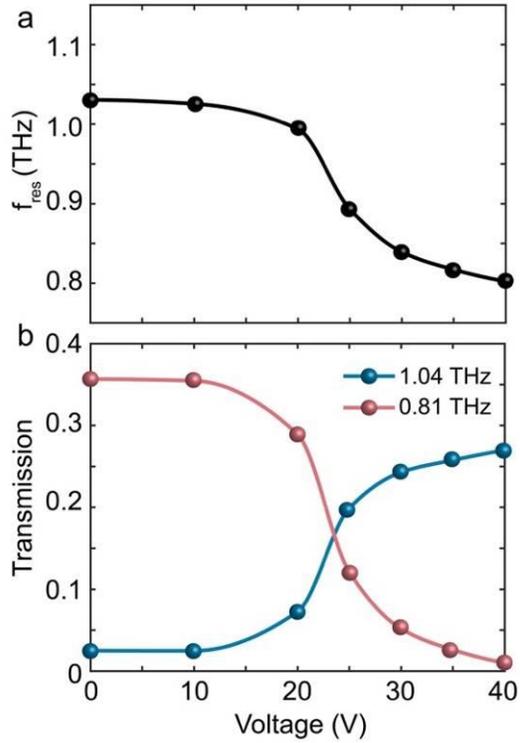

Figure 4. (a) The resonance frequency of x polarization transmission coefficient shifts to lower frequency (redshifts) as the applied voltage increases, showing a 230 GHz tuning range of the resonance frequency. (b) Transmission amplitude for different voltages from 0 V to 40 V at 1.04 THz (blue) and 0.81 THz (red), demonstrating the amplitude modulation capability of the metasurface.

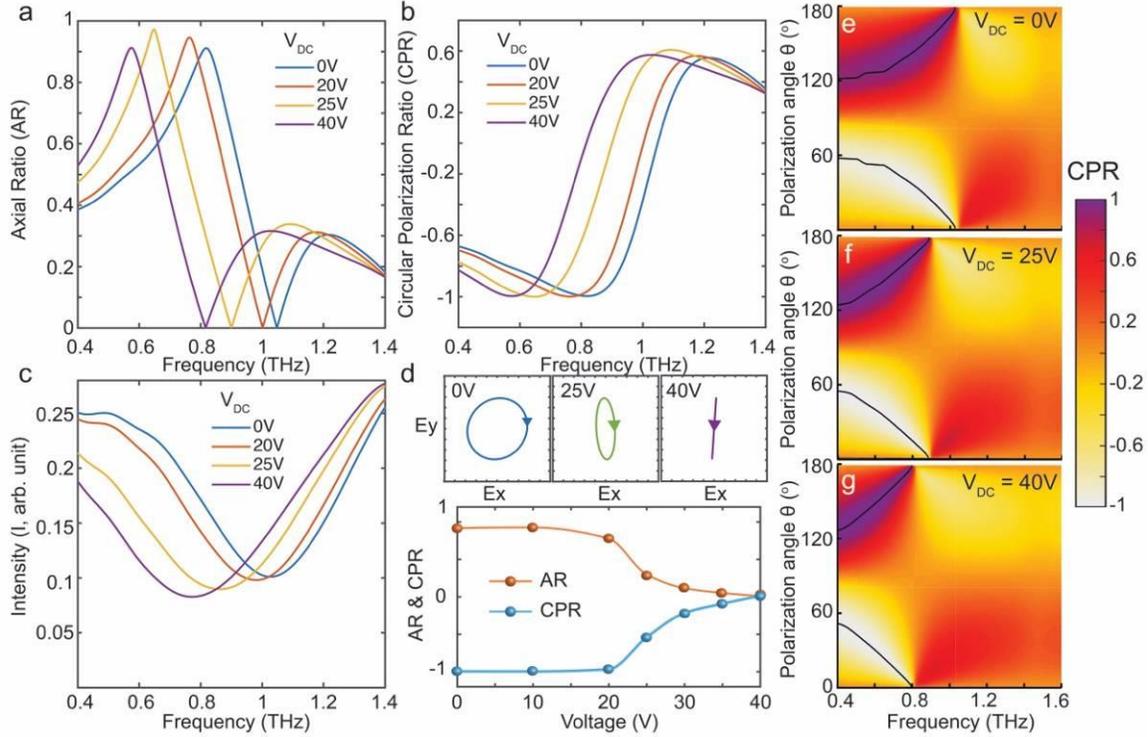

Figure 5. (a) – (c) Experimental axial ratio (a), circular polarization ratio (b), and intensity spectra (c) for various applied voltages describing the polarization state of the transmitted waves, for $\theta = 34°$. (d) Top row is the electric field of the transmitted wave at 0.81 THz for different voltages, demonstrating the capability of tuning the polarization state; the bottom is the AR and CPR at different voltages for $\theta = 34°$. (e) – (g) CPR spectra of transmitted waves for incident waves with different incident polarization angles (from 0° to 180°) at applied voltage of 0 V (e), 25 V (f), and 40 V (g), respectively. The dark lines indicate the frequency closest to pure circular polarization for each incident polarization angle. AR: axial ratio; CPR: circular polarization ratio.

**Supplemental Information: Electromechanically Tunable Metasurface Transmission Waveplate at Terahertz Frequencies**

Xiaoguang Zhao[1], Jacob Schalch[2], Jingdi Zhang[2], Huseyin R. Seren[1], Guangwu Duan[1], Richard D. Averitt[2,*], and Xin Zhang[1,*]

1. Department of Mechanical Engineering, Boston University, Boston, Massachusetts 02215, USA.

2. Department of Physics, University of California, San Diego, La Jolla, California 92093, USA

**1. Fabrication process**

We developed a surface micromachining process to fabricate the tunable metasurface waveplate. First, 400-nm-thick silicon nitride films were coated on both sides of the substrate using low-pressure chemical vapor deposition (LPCVD). Then, photolithography was performed on the top side of the wafer followed by reactive ion etching (RIE) to open windows for ground electrodes. Subsequently, a layer of aluminum was patterned as ground pads using a lift-off process with annealing in $H_2/N_2$ mixing gas at 400 ºC for 30 minutes to obtain ohmic contact between the metal and silicon. Next, ground pads and interconnect wires were patterned by subsequent processes including photolithography, e-beam evaporation of 10-nm chromium and 150-nm gold layers and lift-off. Afterwards, a 400-nm-thick polyimide film was spin-coated on the wafer and cured in $N_2$ ambient at 275 ºC for 1 hour as the sacrificial layer. The cured polyimide film was etched through at the anchors with a titanium layer as the mask and partially etched at the cantilever tips for the formation of dimples. Then, 1-μm copper cantilever structures with 10-nm chromium adhesive films were patterned using the lift-off process. Finally, the sacrificial polyimide layer was completely removed by employing $O_2$ plasma etching.

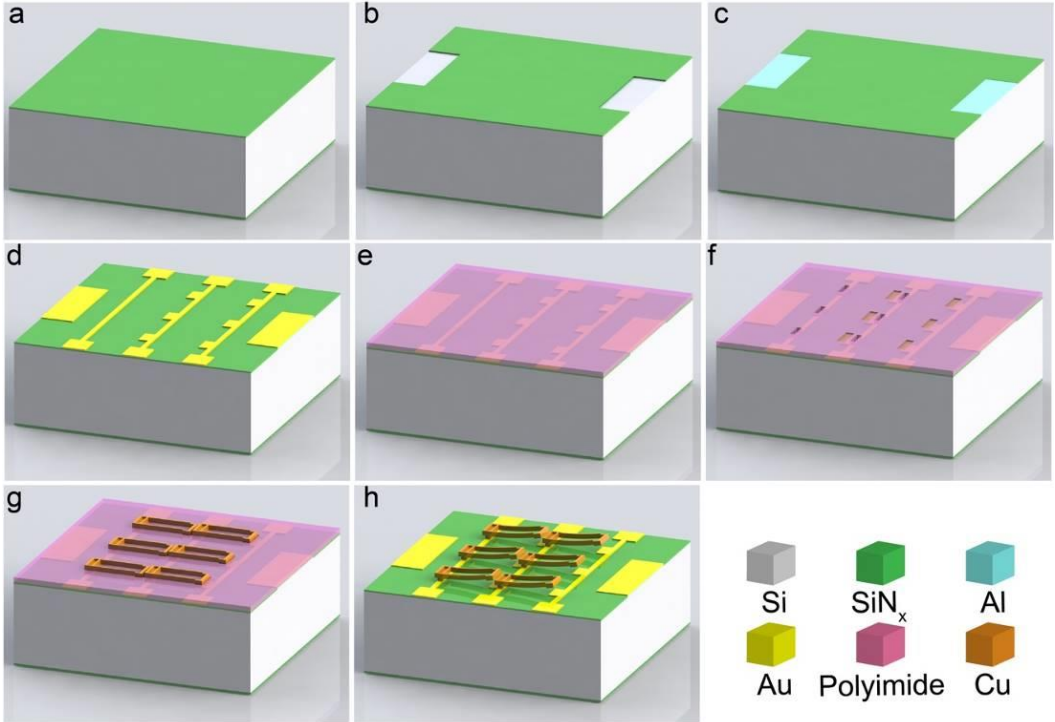

*Figure S1 Fabrication process flow of cantilever tunable metamaterials. (a) Deposit SiN$_x$ layers using LPCVD; (b) Etch opening windows on the top SiN$_x$ layer using RIE; (c) Deposit Al layers for the ground electrodes and anneal at 400 °C in H$_2$/N$_2$ mixing gas for 30 mins; (d) Deposit gold layer serving as ground electrodes and interconnection wires; (e) Spin coat and cure polyimide sacrificial layer; (f) Etch polyimide for anchors of the cantilevers and cantilever dimple; (g) Deposit copper cantilever beam structures; (h) Release the cantilevers by etching the polyimide sacrificial layer completely. Details of the process parameters can be found in the Methods section of the main text and elsewhere [ref.S1].*

## 2. Terahertz time domain spectroscopy based on photo-conductive antennas

The fabricated samples were characterized using the terahertz time domain spectroscopy (THz-TDS) based on photo-conductive antennas, as shown in Fig. S2a. An 800 nm Ti:Sapphire laser with 25 fs pulse duration and 80 MHz repetition rate is split by a beam splitter to form two beams, one of which (generation beam) is employed to generate THz pulses while the other (detection beam) is for detection. The generation beam impinges on a biased photo-conductive antenna through a set of optics to generate the THz pulses, which are focused on the sample using a pair of off-axis parabolic mirrors. The polarization angle of the incident THz pulses relative to the cantilever beams can be adjusted by rotating the metasurface sample about the incident axis of the pulses. The transmitted THz pulses are focused on the detector photo-conductive antenna, which generates a current that is proportional to the instantaneous THz electric field strength. The time resolved field strength of transmitted waves is obtained by controlling the time delay between the THz pulses and the detection beam. The transmission through air is used as the reference and the temporal transmission response for x-polarized incident pulses are recorded for each applied driving voltage, as shown in Fig. S2b. Then, we Fourier transform the time domain response and normalize the frequency response with the reference to obtain the metasurface spectrum (Fig. S2c). The time domain response is windowed to crop the multi-reflections (shaded area in Fig. S2b) caused by the 300-μm-thick silicon substrate. We can eliminate the multi-reflection effect by thinning the substrate to a few microns. The response to the y-polarized THz pulses are obtained by rotating the sample by 90°.

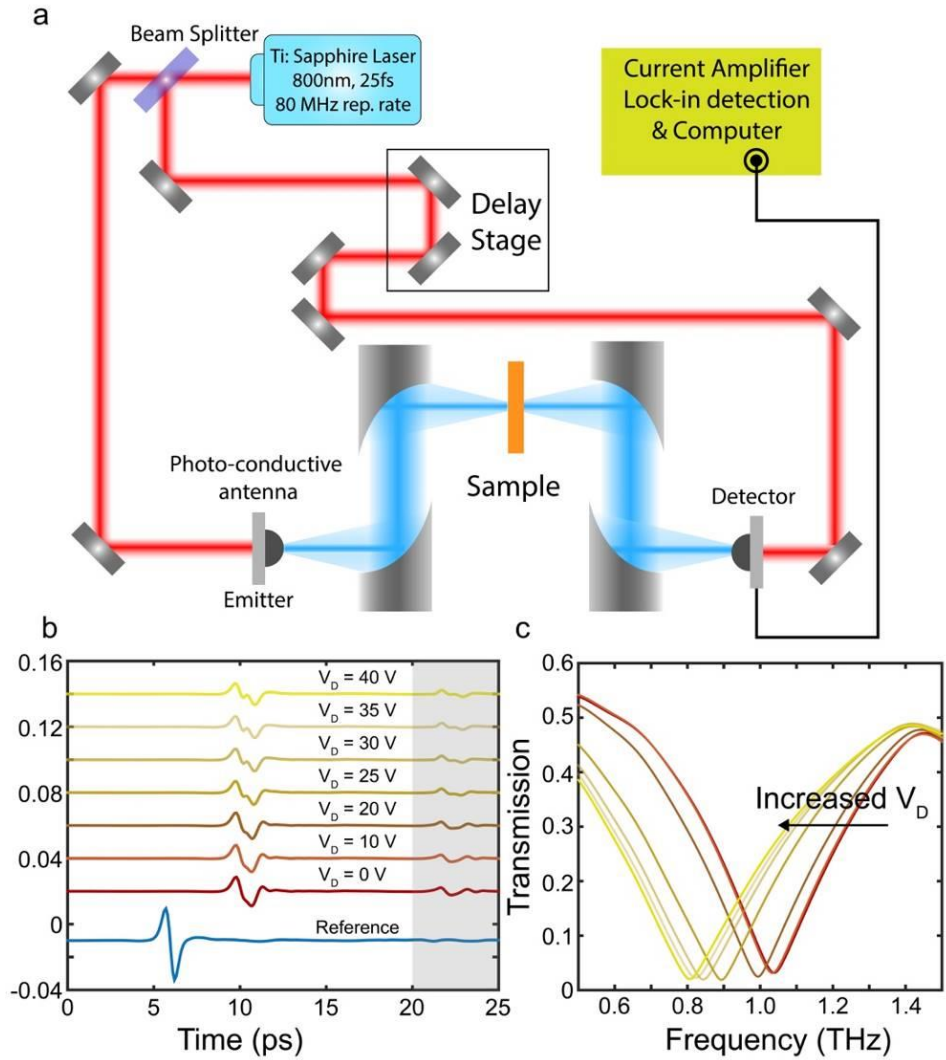

*Figure S2. (a) The THz time domain spectroscopy system based on photo-conductive antennas. (b) The time domain curves of the reference and the metasurface under different applied driving voltages. The shaded area represents the reflections caused by the substrate, which is windowed to remove etalon-based oscillations in the frequency domain. (c) The calculated spectrum of the metasurface under different applied voltages.*

## 3. Equivalent circuit model of the tunable metasurface

The transmission coefficient of the metasurface can be intuitively explained using the equivalent circuit model. As shown in Figs. 1b and 1c, the response to x polarized light is equivalent to a RLC resonator while the response to y polarized light is equivalent to a RL circuit. According to transmission line (T.L.) theory, the transmission coefficients for x and y polarized incidence, denoted by $t_{xx}$ and $t_{yy}$, respectively, can be calculated by Eqs. (2) and (3) in the main text. In the equations, $Z_{xx} = R_1 + i\omega L_1 + 1/(i\omega C_1)$ and $Z_{yy} = R_2 + i\omega L_2$, where $R_i$ ($i$ = 1, 2) is the resistance, $L_i$ ($i$ = 1, 2) is the inductance, and $C_1$ is the capacitance, as shown in the circuit model in Figs. 1b and 1c. Thus, the transmission coefficients can be written as

$$t_{xx} = \frac{2[R_1 + i\omega L_1 + 1/(i\omega C_1)]}{R_1 + i\omega L_1 + 1/(i\omega C_1) + Z_0} \cdot D \quad \text{(S.1)}$$

$$t_{yy} = \frac{2(R_2 + i\omega L_2)}{R_2 + i\omega L_2 + Z_0} \cdot D \quad \text{(S.2)}$$

where $D$ accounts for the transmission loss at the substrate/air interface, and $Z_0$ is the free space impedance (377 Ω). Theoretically, the resistance, inductance, and capacitance can be calculated; however, parasitic effects in the sub-wavelength resonators make it difficult to obtain accurate analytical models for these values. Instead, we applied the least square fitting to retrieve the unknown variables ($R_i$, $L_i$, $C_1$, and D) in the above equations. The simulation results were used in the fitting due to 1) that the simulation results agree with the experiments (as shown in Fig. 2 in the main text) and 2) that they exhibit fewer experimental artifacts in the spectrum. The experimental results, simulation results and transmission line model fitting curves for each voltage are shown in Fig. S3, and exhibit excellent agreement, validating the analytical model. The retrieved variables are listed in Table S1. In the fitting results, the majority of the variables, including $R_1$, $L_1$, $R_2$, $L_2$, and $D$, are nearly independent of the applied voltage, while $C_1$ increases as the applied voltage increases. The results manifest the physics underlying the tunability. For the x polarized incidence, the applied voltage pulls the cantilever downwards to increase the capacitance between the cantilever tip and the underlying capacitive pad, with very little effect on the inductance or resistance of the cantilevers. For the y polarized light, the deformation of the cantilever does not affect the equivalent parameters.

In the fabricated devices, the variation of the capacitance ($C_1$) is limited since the deformation is limited by the pull-in effect. Larger changes in the capacitance could be achieved through fabrication by, for example, eliminating the oxidation of cantilevers and decreasing the roughness of cantilever tips. We can predict $t_{xx}$ of the metasurface for a larger tuning range of $C_1$, as shown in Fig. S4. If it was possible to tune $C_1$ from 0.2 fF to 1.2 fF, the resonant frequency would be shifted from 1.4 THz to 0.6 THz. If ohmic contact could be established (with further fabrication process optimization), then when cantilever tips touch capacitive pads, the resonance will be eliminated since the capacitance would be shunted. Moreover, the analytical model of the transmission coefficients can be used to design the lumped parameters for other working frequencies and optimize polarization control as discussed below.

*Table S1. The retrieved variables in the equivalent circuit model for different applied voltages.*

| $V_{DC}$ (V) | $R_1$ (Ω) | $L_1$ (pH) | $C_1$ (fF) | $R_2$ (Ω) | $L_2$ (pH) | D |
|---|---|---|---|---|---|---|
| 0 | 4.9 | 69.5 | 0.32 | 61.1 | 93.7 | 0.31 |
| 20 | 5.6 | 66.6 | 0.38 | 61.1 | 93.7 | 0.31 |
| 25 | 4.9 | 61.4 | 0.52 | 64.8 | 90.7 | 0.31 |
| 30 | 4.3 | 60.0 | 0.60 | 64.8 | 90.8 | 0.31 |
| 35 | 4.8 | 58.3 | 0.67 | 61.7 | 92.0 | 0.31 |
| 40 | 4.7 | 55.4 | 0.69 | 61.7 | 92.0 | 0.31 |

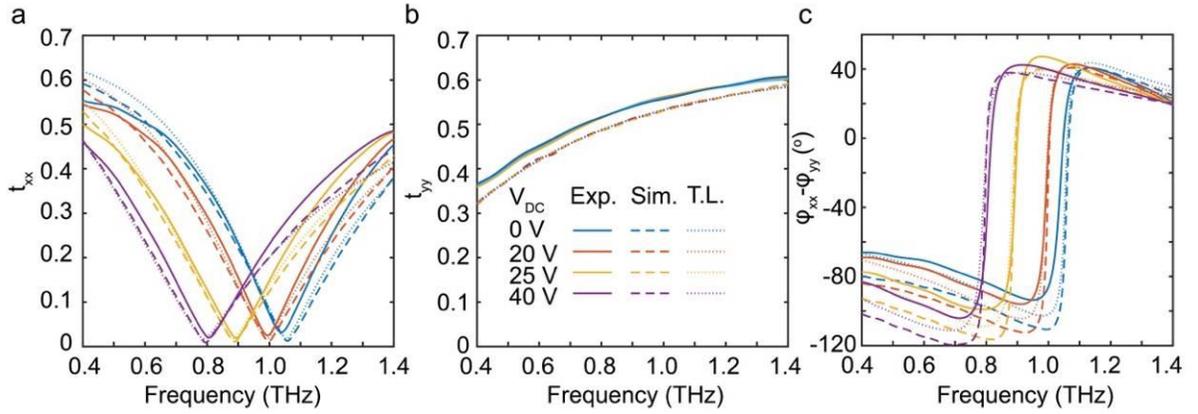

*Figure S3. The experimental (solid lines), simulation (dash lines), and transmission line modeled (dotted lines) transmission coefficients of the metasurface, including: transmission amplitude for (a) x polarization incidence ($|t_{xx}|$) and for (b) y polarized incidence ($|t_{yy}|$), and (c) the phase difference between $t_{xx}$ and $t_{yy}$, i.e. ($\varphi_{xx}$-$\varphi_{yy}$).*

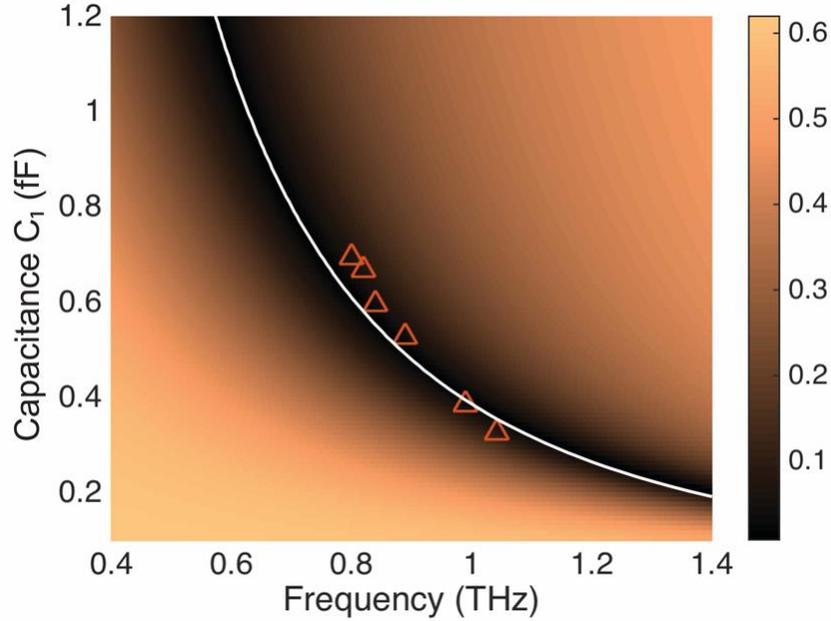

*Figure S4. The calculated results of the transmission coefficients ($t_{xx}$) upon varying $C_1$ by using the retrieved parameters in Table S1, i.e. $L_1 = 65$ pH and $R_1 = 5\ \Omega$. The white solid line is a guide to the eye for the resonant frequency. The red triangles (Δ) represent the experimentally measured resonant frequencies for different voltages.*

### 4. Tunable polarization state revealed by Stokes parameters

Stokes parameters of the transmitted waves can be derived using the equations in the Methods section of the main text. The full set of stokes parameters, including $S_0$, $S_1$, $S_2$, and $S_3$, can be plotted in a Poincare plot to visualize the polarization state of a wave. Fig. S5a is the Stokes parameters for 34° incident polarization angle, calculated from experimental transmission coefficients and plotted on a normalized Poincare sphere. Each point on the curves corresponds to the Stokes parameters of a specific frequency and each curve corresponds to one specific applied voltage. In the Poincare sphere, the point intersects with the z-axis (i.e., $S_1 = 0$, $S_2 = 0$, and $S_3 = \pm 1$) corresponds to pure circular polarization while the intersection with the x-y plane (i.e. $S_1 \neq 0$, $S_2 \neq 0$, and $S_3 = 0$) corresponds to linear polarization. All other points represent elliptical polarization states. We can read the orientation angle (ψ) and ellipticity angle (χ), which are depicted in Fig. S5b, directly from the plot. In Fig. S5a, the Stokes parameters at 0.81 THz are highlighted by the solid dots for different applied voltages. When $V_{DC}$ = 0 V, it is close to (0, 0, -1), manifesting circular polarization. When $V_{DC}$ = 40 V, the Stokes parameters locate at (-0.99, 0.13, 0), meaning linear polarization. The polarization states are plotted in Fig. S5c for each voltage. We can capture more information from the full set of Stokes parameters using Poincare sphere.

Fig. S6 shows the tunable polarization states of transmitted waves for different incident angles at different frequencies, corresponding to the most nearly circular polarized transmission. For example, the transmitted wave is right-hand circularly polarized for incident polarization angle of 26° at 0.88 THz when the applied voltage is 0 V. An increase in the voltage (from 0 V to 25 V) gradually tunes the polarization state to elliptical polarization. When the applied voltage is 40 V, the transmitted wave is left-hand elliptically polarized. It is similar for other incident polarization angles.

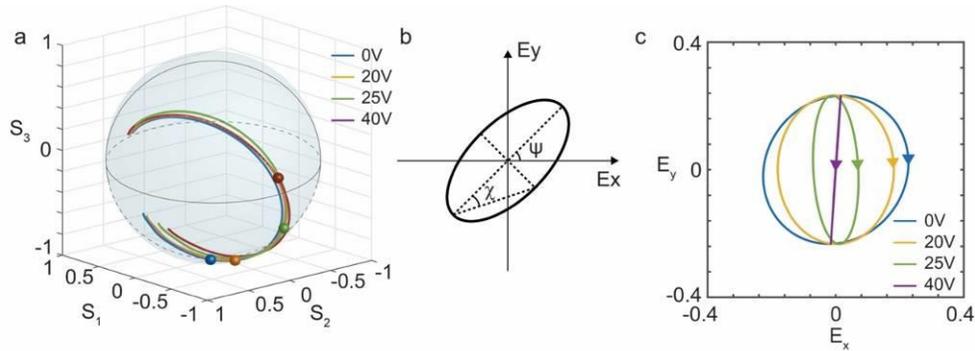

*Figure S5. (a). Poincare plot of the transmitted waves for different applied voltage in the frequency range from 0.4 THz to 1.4 THz. With increasing frequency, the corresponding Stokes parameters shift from the bottom left to upper left along the sphere. The solid dots correspond to 0.81 THz. (b). Illustration of an elliptically polarized wave to define orientation angle (ψ) and*

*ellipticity angle (χ). (c). The electric field components along x- and y-axis of the transmitted waves to illustrate the polarization states at 0.88 THz.*

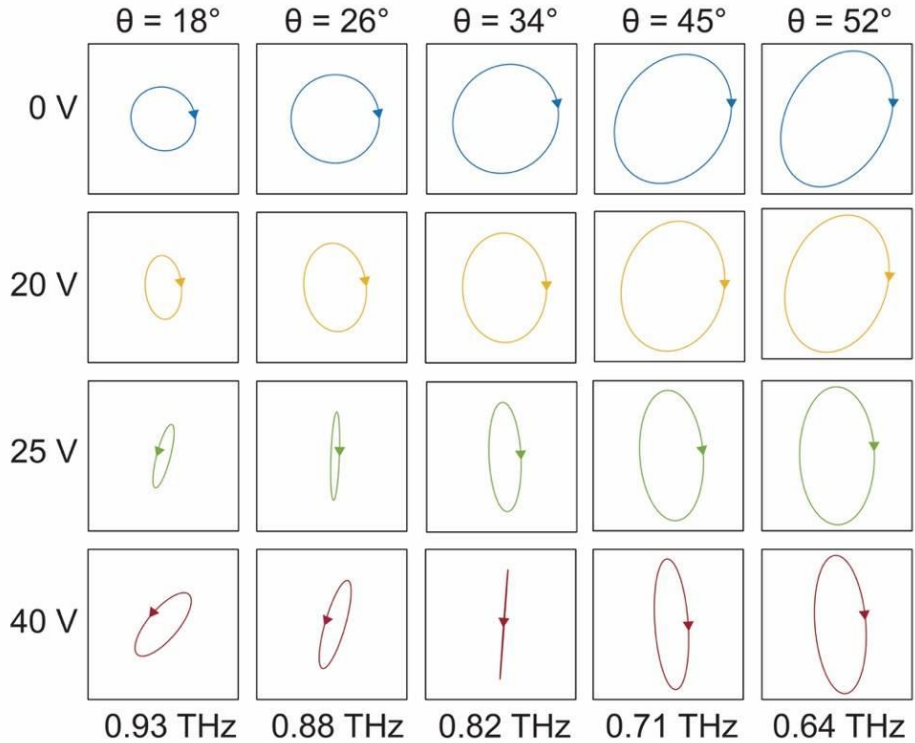

*Figure S6. The polarization state for different incident polarization angles - θ - (and frequencies) and applied voltages.*

**References:**

[S1]. X. Zhao, J. Schalch, J. Zhang, H.R Seren, G. Duan, R.D. Averitt, and X. Zhang, "A tunable terahertz metamaterial based on a micro-cantilever array," 2017 IEEE 30th International conference on Micro Electro Mechanical Systems (MEMS), Las Vegas, NV, 2017, pp. 910-973.